\documentclass[proof]{WileyASNA-v1}

\articletype{Article Type}%
\usepackage{amssymb}

\begin{document}

\title{Formation of spiral structure from the violent relaxation of self-gravitating disks}

\author[sut,rmutsb]{T. Worrakitpoonpon*}

\authormark{Worrakitpoonpon}

\address[sut]{School of Physics, Institute of Science, Suranaree University of
  Technology, Nakhon Ratchasima 30000, Thailand}
\address[rmutsb]{Faculty of Science and Technology, Rajamangala University of
  Technology Suvarnabhumi, Nonthaburi 11000, Thailand}

\corres{Tirawut Worrakitpoonpon, Suranaree University of Technology,
  Nakhon Ratchasima 30000, Thailand \email{worraki@gmail.com}}

\abstract{
  We present the numerical study of the formation of spiral structure in 
  the context of violent relaxation. Initial conditions are the
  out-of-equilibrium disks of self-gravitating particles in rigid rotation.
  By that mechanism, robust and non-stationary spiral arms can be formed
  within a few free-fall times by the shearing of the mass ejection
  following the collapse. With a closer look, we find different properties
  of the arms in connection with the initial configuration. The winding
  degree tends to increase with initial angular speed provided that a
  disk is thin. If disk surface is circular, both number and position of
  arms are governed by the Poissonian density fluctuations that produce
  more arms as more particles are introduced.
  On the contrary, if the surface ellipticity is imposed, the number of arms
  and their placement are effectively controlled. Otherwise, the increase of
  thickness leads to a complicated outcome since the number of arms and winding
  degree are less effectively controlled. We speculate that this complexity is caused
  by a strong non-axisymmetric field during the violent relaxation that 
  disorganizes the pre-collapse motion and the concentration of particles.
}


\maketitle

\section{Introduction} \label{intro}

The fact that a system governed by self-gravity, or more generally
the long-range interaction, starting far from equilibrium
undergoes the violent relaxation leading to the quasi-stationary state (QSS)
has been established as a standard paradigm in the field of non-equilibrium
statistical mechanics. This theory has been proposed by \citet{lynden_bell_1967}
in attempt to resolve the deviation of the observed light distribution of elliptical
galaxies from the isothermal profile. His work demonstrates that, under the
time-varying mean-field created during this stage, the system relaxes
to a predictable QSS within a time-scale of order dynamical times.
In the following years, the discrepancy arises because numerous simulations 
reveal that there does not exist the universal QSS as proposed
by the Lynden-Bell theory. The relaxed states exhibit
great diversities of density profiles and the velocity distributions
and almost all of them can not be described by the Lynden-Bell statistics
(see, e.g., \citealt{aguilar+merritt_1990, roy+perez_2004, trenti_et_al_2005,
  boily+athanassoula_2006, barnes+lanzel+williams_2009, sylos_labini_2013}).
Only when the relaxation is not too violent, the Lynden-Bell
distribution fits the QSS \citep{levin+pakter+rizzato_2008}.
In addition, the violent relaxation of a rotating spheroid
has also been examined to explain the highly flattened configuration of the
elliptical galaxies \citep{gott_1973, aguilar+merritt_1990, katz_1991}.

While the majority of studies addresses principally to the ellipticity
and the density profile of the QSS, a recent work 
by \citet{benhaiem_et_al_2017} proposes that the formation of long-lived
non-stationary spiral structure by the violent relaxation is possible
if one starts from gravitationally unstable rotating ellipsoids.
This provides an alternative
scenario different from the standard framework of this issue
which believes that the spiral pattern is the perturbed
component of the disk or bulge in equilibrium
\citep{lin+shu_1964, goldreich+lynden_bell_1965b} or the tidal tails
by the passage of a companion object \citep{toomre+toomre_1972}.
Following work of the same group explores further  
and demonstrates that some specific structures such as bar and ring
are also obtainable by the same mechanism \citep{benhaiem_et_al_2019}.
In this work, we present the numerical study of the
formation of spiral pattern by the violent relaxation starting from
an out-of-equilibrium rotating system with rigid rotation.
We simplify the initial shape to be a disk in order to conveniently
investigate the formation process, which is confined onto the disk plane.
Main purposes are to understand theoretically the physical process that
generates the spiral arms in this simple model and how the arm properties
depend on the initial disk parameters. While this formation scenario is
different from the standard framework: the spiral arms arise as perturbations
in a differentially rotating disk in equilibrium, we will show that
many underlying processes forming and evolving our spiral structure
shares similarities with those in the astrophysical disk.
The article is organized as follows. First in Sec. \ref{eq_motion}, we briefly 
discuss the equations of motion of the corresponding system and how it could 
lead to the final spiral pattern. In Sec. \ref{nume_para}, we introduce the 
initial conditions, the simulation set-up, the accuracy control 
and the necessary parameters for measurement.
In the following section, we present the numerical results,
including mainly the different morphologies of the spiral structures in each
disk type as well as their underlying mechanisms.
Finally in Sec. \ref{discussion}, we provide the conclusion and discussion.

\section{Equation of motion for collapsing disk in rotation}
\label{eq_motion}

Let us consider a self-gravitating planar disk with circular shape and
uniform surface density. It is put in a rigid rotation with 
angular speed $\omega_{0}$, well below that required to prevent the collapse,
without the velocity dispersion. At the starting time,
the radial equation of motion of a test particle located at radius $r_{0}$ 
in the rotating polar coordinate in coherence with $\omega_{0}$ reads
\begin{equation}
  \frac{d^{2}r}{dt^{2}}-r\bigg( \frac{d\theta}{dt}\bigg)^{2} = 
  \omega_{0}^{2}r+2\omega_{0}r\frac{d\theta}{dt}-\nabla_{r}\phi (r)\Big|_{r=r_{0}}
  \label{eq_mot_disk_rad}
\end{equation}
where the first and the second terms on the right-hand side stand for
the fictive centrifugal and radial Coriolis forces, respectively.  
The third term involves the gravitational potential.
Semi-analytical expression of the potential inside a uniform circular disk
at radius $r$ governed by inverse-distance mutual potential
was proposed by \citet{ciftja+hysi_2011} to be
\begin{equation}
  \phi (r) = \frac{2\phi_{0}}{\pi} E\Bigg[ \bigg(\frac{r}{r_{d}}\bigg)^{2} \Bigg]
  \label{poten_disk}
\end{equation}
where $\phi_{0}$ is the potential at center and $r_{d}$ is the disk
radial size. The function $E$ is given by
\begin{equation}
  E(m) = \int_{0}^{\pi /2}d\theta \sqrt{1-m\sin^{2}\theta} \label{bessel_eq}
\end{equation}
for $m\leq 1$. We re-scale the radius in equation (\ref{eq_mot_disk_rad})
to be $R \equiv r/r_{0}$ and the equation then reads
\begin{equation}
  \frac{d^{2}R}{dt^{2}}-R\bigg( \frac{d\theta}{dt}\bigg)^{2} = 
  \omega_{0}^{2}R+2\omega_{0}R\frac{d\theta}{dt}-
  \nabla_{R}\tilde{\phi} (R;r_{0})\Big|_{R=1}
  \label{eq_mot_disk_rad_r}
\end{equation}
where $r_{0}$ is merged into the new expression of potential $\tilde{\phi}$.
Similarly, the tangential equation of motion in terms of $R$ reads
\begin{equation}
  R\frac{d^{2}\theta}{dt^{2}}+2\frac{dR}{dt}\cdot\frac{d\theta}{dt} =
  -2\omega_{0}\frac{dR}{dt}
  \label{eq_mot_disk_tan_r}
\end{equation}
where the right-hand side corresponds to the tangential Coriolis force.

As we have seen in equations (\ref{eq_mot_disk_rad_r}) and (\ref{eq_mot_disk_tan_r}),
the introduction of $\omega_{0}$ makes the evolutions
of $R$ and $\theta$ not depend only on time $t$ but $r_{0}$ is also
involved in the particle motion via the gravitational potential at start.
In other words, the memory of the initial position of each particle is retained
along the dynamics. We then re-arrange equation (\ref{eq_mot_disk_tan_r})
and express it back in terms of $r$ as
\begin{equation}
  \frac{d\dot{\theta}}{d(\ln r)}\sim -\omega_{0}. 
  \label{eq_mot_disk_dthetadotdr}
\end{equation}
This expression can be interpreted by that 
the shearing strength, i.e. the departure from the rotating
frame in a differential way, is fuelled by $\omega_{0}$ itself.
These are the key features that we capture and will inspect how
they can produce the spiral structure in the $N$-body simulations.

\section{Preparation of simulations and parameters} \label{nume_para}
  
\subsection{Initial conditions and choice of units} \label{ic}

The initial condition is the disk with uniform thickness $z_{0}$
and density $\rho_{0}$. Disk cross-section is characterized
by the ellipticity defined as $e_{0}\equiv R_{a}/R_{b}-1$ where $R_{a}$
and $R_{b}$ are semi-major and -minor axis lengths, respectively.
The disk scale-length is that $R_{b}$ is fixed to $1$. 
The typical number of particles $N$ is $128,000$, 
if not indicated otherwise. To generate the initial configuration,
particles are thrown randomly into a confining space of a given size.
The initial motion of particles is purely rotational with constant
angular speed $\omega_{0}$ regardless of position which is
numerically adjusted to satisfy the initial virial ratio
$b_{0}\equiv 2T_{0}/|U_{0}|$ where $T_{0}$ and $U_{0}$ are the initial
kinetic and potential energies, respectively.
Time unit is 
\begin{equation}
  t_{d} = \sqrt{\frac{1}{G\rho_{0}}}   \label{def_td}
\end{equation}
where $G$ is the Newton's gravitational constant.
The cases we have simulated are summarized in Tab. \ref{tab_disk} 
with their code names. From that table, we can divide those cases
into three groups: TC=thin-circular, TN=thin-elliptical
and TK=thick-elliptical. Segment of the code name, e.g. TN, E01 or B03, 
may be employed to represent sub-group of those initial conditions 
that we will address to.

\begin{table}
  \caption{Summary of the simulated cases with their physical
    parameters and code names. Note that TC=thin-circular,
    TN=thin-elliptical and TK=thick-elliptical.}
  \begin{center}
    \begin{tabular}{c|c|c|c}
      \hline \hline
      Code & $e_{0}$ & $z_{0} $ & $b_{0}$ \\
      \hline \hline
      TC-B01 & $0.0$ & $0.1$ & $0.1$ \\
      TC-B02 & $0.0$ & $0.1$ & $0.2$ \\
      TC-B03 & $0.0$ & $0.1$ & $0.3$ \\
      \hline
      TN-E01-B01 & $0.1$ & $0.1$ & $0.1$ \\
      TN-E01-B02 & $0.1$ & $0.1$ & $0.2$ \\
      TN-E01-B03 & $0.1$ & $0.1$ & $0.3$ \\
      TN-E01-B04 & $0.1$ & $0.1$ & $0.4$ \\
      \hline
      TN-E02-B01 & $0.2$ & $0.1$ & $0.1$ \\
      TN-E02-B02 & $0.2$ & $0.1$ & $0.2$ \\
      TN-E02-B03 & $0.2$ & $0.1$ & $0.3$ \\
      TN-E02-B04 & $0.2$ & $0.1$ & $0.4$ \\
      \hline
      TK-E01-B01 & $0.1$ & $0.4$ & $0.1$ \\
      TK-E01-B02 & $0.1$ & $0.4$ & $0.2$ \\
      TK-E01-B03 & $0.1$ & $0.4$ & $0.3$ \\
      TK-E01-B04 & $0.1$ & $0.4$ & $0.4$ \\
      \hline
      TK-E02-B01 & $0.2$ & $0.4$ & $0.1$ \\
      TK-E02-B02 & $0.2$ & $0.4$ & $0.2$ \\
      TK-E02-B03 & $0.2$ & $0.4$ & $0.3$ \\
      TK-E02-B04 & $0.2$ & $0.4$ & $0.4$ \\
      \hline \hline
    \end{tabular}
  \end{center}
\label{tab_disk}
\end{table}

\subsection{Simulation set-up and control} \label{nume_setup}
The dynamics of particles in the Newtonian space without an expansion
of background is handled by
GADGET-2 \citep{springel+yoshida+white_2001, springel_2005}.
In the code, gravitational force is softened by a spline kernel. 
Below a pre-defined length, the force converges to zero at zero separation.
Above the same scale, the force is Newtonian.
The softening length in all simulations
is fixed to $R_{b}/720$. The integral time-step is fixed to
$t_{d}/12,000$ for TC cases where the dynamics is sensitive to the
density fluctuations. For TN and TK disks, it is controlled
to be from $t_{d}/6,000$ to $t_{d}/4,000$ until the typical
ending time. In the prolonged stage of some selected TN, it is 
from $t_{d}/4,000$ to $t_{d}/3,000$.
With this control, the deviation of the total energy at 
any time is not more than $0.1\%$ for TC and TN and $0.2\%$ for TK. 
It is controlled below $1\%$ in the prolonged TN runs.
The total angular momentum is conserved within $0.01\%$ of
deviation in the entire simulation.

\subsection{Physical parameters for measurement} \label{pitch}

In addition to the standard parameters that we will employ,
we introduce a few more parameters that might be useful.
The first one is
the parameter to quantify the winding degree of the spiral arms.
We choose the pitch angle extracted from
the logarithmic spiral function
which is extensively used to study the spiral galaxies.
With a constant pitch angle, we can re-arrange 
it in polar coordinate $(r,\theta )$ to be
\begin{equation}
  r(\theta ) = r_{s}e^{\theta \tan \mu} \label{pitch_def}
\end{equation}
where $r_{s}$ is a constant and $\mu$ is the pitch angle.
By definition, $|\mu|$ ranges from $0^{\circ}$ 
to $90^{\circ}$, i.e. from circumferential to radial contour.
If $\mu$ is positive, the spiral is `leading'. 
The opposite sign is for the `trailing' spiral arm.

The second quantity is the collapse factor $\mathcal{C}$ defined as 
\begin{equation}
  \mathcal{C}=\frac{U(t)}{U(0)} \label{col_def}
\end{equation}
where $U(t)$ and $U(0)$ are the system potential energies at time $t$ and at
the initial time, respectively.
This parameter measures the compactness of a
system during the evolution by mean of comparing the depth of 
the potential energy relative to the initial value.

Third, we introduce the ellipticity parameter of
the disk surface defined as 
\begin{equation}
  \iota = \frac{\Lambda_{2}}{\Lambda_{1}}-1 \label{iota_def}
\end{equation}
where $\Lambda_{2}$ and $\Lambda_{1}$ correspond to the higher
and lower moments
of inertia of the principal axes on the disk plane, respectively. These values
can be obtained by solving the eigenvalue equation of an inertia tensor.

\section{Numerical results} \label{nume}

\subsection{Kinematics of the formation of spiral arms for TC disk} 
\label{early_winding}

\begin{figure*}
  \begin{center}
    \begin{tabular}{ccc}
      \hspace{-8mm}
      \includegraphics[width=5.6cm]{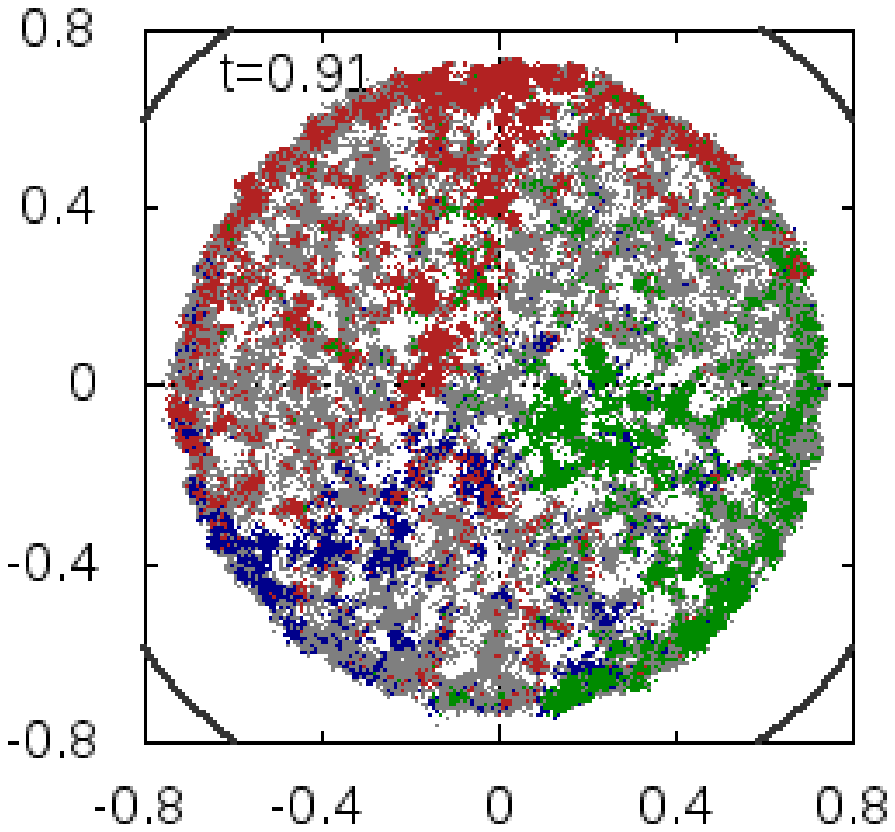}  \hspace{-5mm}
      & \includegraphics[width=5.6cm]{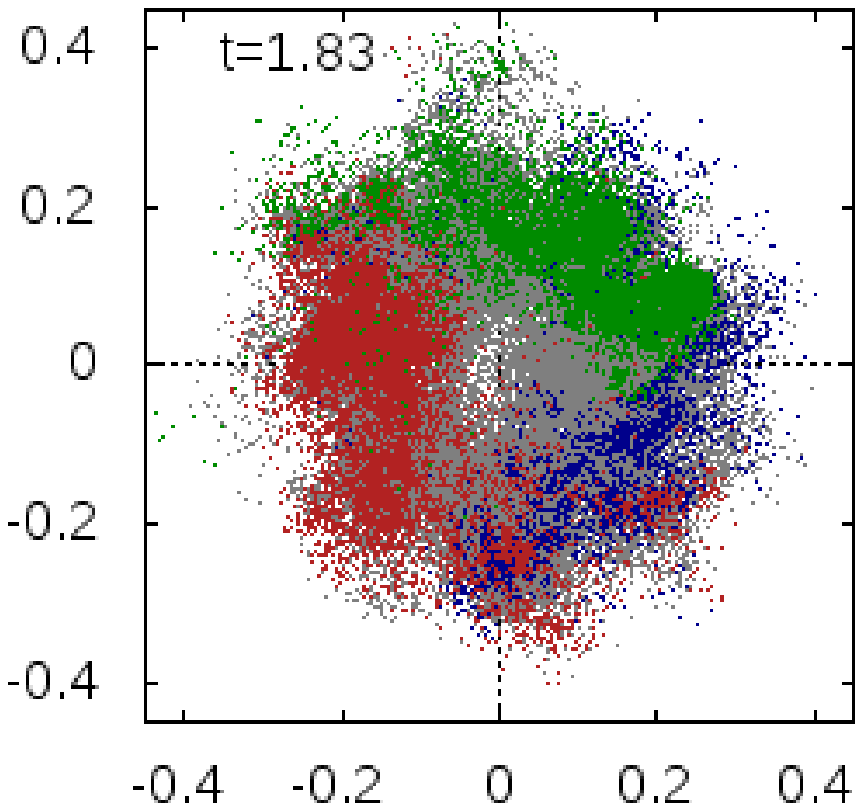} \hspace{-5mm}
      & \includegraphics[width=5.6cm]{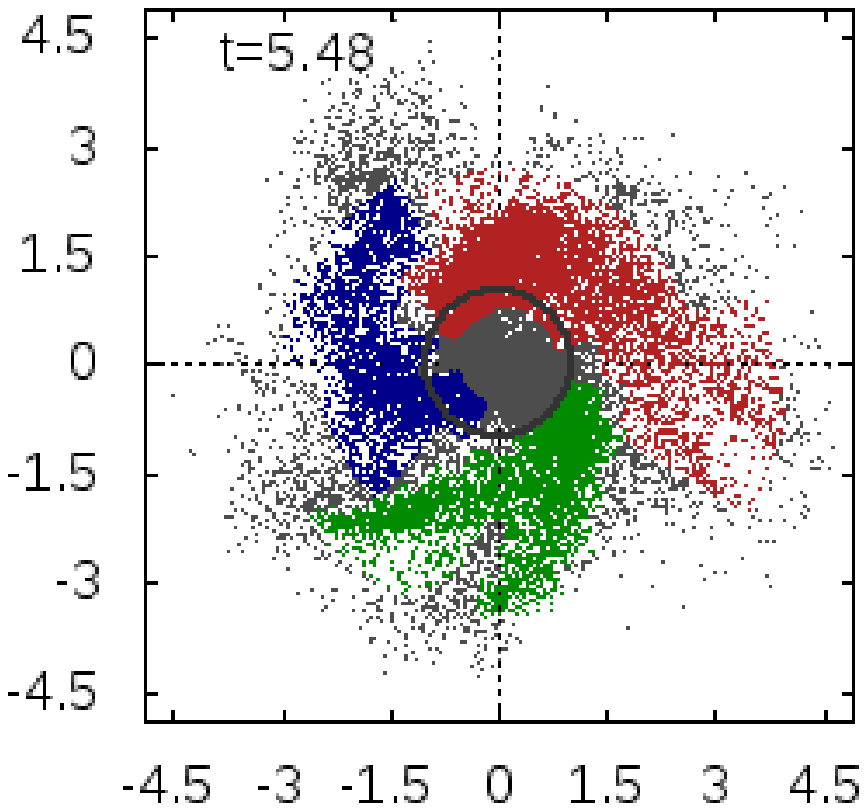}  
    \end{tabular}
  \end{center}
  \caption{The configuration of TC-B02 projected onto the disk plane
    at different indicated times. Particles constituting each arm observed at
    $5.48 \ t_{d}$ are distinguished by different colors. The size of
    the initial disk is depicted by a solid line.
    This plot underlines the importance of the Poissonian
    density fluctuations to form the spiral arms in a TC disk.}
  \label{fig_circular_early}
\end{figure*}

First of all, we visually examine the formation of the spiral pattern in 
a TC disk. Shown in Fig. \ref{fig_circular_early} is the configuration of
TC-B02 projected onto the disk plane at different times around the maximum
contraction (see information of the presentation in caption).  
At the final time, we observe the trailing spiral arms emerging from
three different points of the central nucleus. With a close inspection
on the evolution detail, we observe at $0.91 \ t_{d}$ the densely-packed
clumps of particles constituting the peripheral ring while the disk
is collapsing. In the inner part, we observe the loosely-distributed
nodes of particles attracted towards the local over-density 
by finite-$N$ fluctuations. There is not yet any trace of
the spiral pattern at this moment but we observe a clear partition
between groups of particles that eventually become each arm.
At $1.83 \ t_{d}$, the disk reaches the minimum size and each population
contracts further, forming dense separated clouds at different
locations. At $5.48 \ t_{d}$, the continuous spiral arms emerge out around
the dense nucleus and they are dwelling in the background with
diluted density, i.e. the inter-arm region.
The system size is rapidly increasing after the collapse.

We may underline from this result that although the point-wise initial
condition has been drawn from an axisymmetric mass distribution function,
these emerging arms are the outcome of the breaking of isotropy
by the finite-$N$ effect. It triggers the separated mass
concentrations and outflows before and after the collapse.

\begin{figure*}
  \begin{center}
    \begin{tabular}{cc}
      \hspace{-6mm}
      \includegraphics[width=6.5cm]{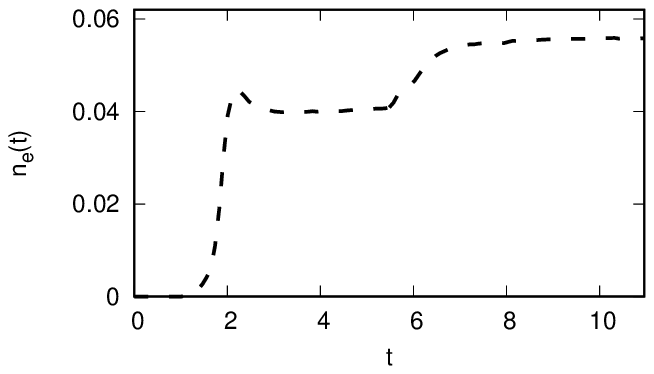} 
    & \includegraphics[width=6.5cm]{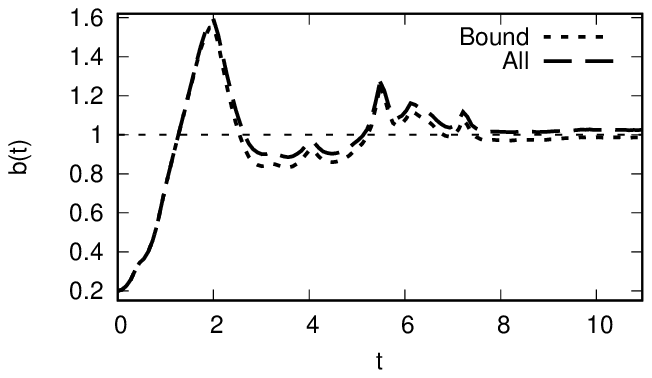}
    \end{tabular}
  \end{center}
  \caption{The fraction of ejected particles $n_{e}$ (left panel)
    and the virial ratio $b$ calculated from bound and all particles
    (right panel) as functions of time for TC-B02.
    It turns out that the violent relaxation of a rotating
    system leads to multiple ejections before it is virialized.}
  \label{fig_cr_ne}
\end{figure*}

We now inspect the procedure of violent relaxation in detail.
Shown in Fig. \ref{fig_cr_ne} are the temporal evolutions
of the fraction of ejected particles $n_{e}$ and the virial ratio $b$
calculated from all and bound particles of the same case.
The ejected (or unbound) particles are those with positive energy
while the bound particles correspond to those with negative energy.
Considering $n_{e}$,
we observe that the ejection does not occur once but there are two events of it.
The first one by gravitational
collapse occurs around $1 \ t_{d}$ and it brings the fraction to $\sim 0.04$.
We also remark that this ejection is not punctual
but it lasts for $\sim 1 \ t_{d}$. Afterwards, this fraction remains
constant until the second ejection when $n_{e}$ is boosted again during
$6-7 \ t_{d}$ and reaches $\sim 0.055$ that is apparently the terminal value.
This can be interpreted by that the particles at different initial positions
are designated to escape at different times. In other words, the time at
which each particle is ejected is $r_{0}$-dependent. The evolution of
the virial ratio of bound structure is in coherence with $n_{e}$:
it increases rapidly during the first collapse before it stays mildly
sub-virialized for a period of time. Then, the secondary ejection brings it up
again and relaxes down to the virialized state.
On the contrary, the virial ratio of the entire structure always remains slightly
above the bound one since the first ejection. This is because the ejected
mass carries the excessive kinetic energy.
These plots inform that, although the spiral structure in
Fig. \ref{fig_circular_early} appears to be fully formed at $5.48 \ t_{d}$,
there is still an ongoing evolution. However, we do not expect any
major effect on the spiral pattern by the second wave of ejection as
it occurs far inside and later in time.
From these results, it turns out that the virialization of a rotating
disk is delayed as it takes more than three times the collapse
time because of the strong collective motion and multiple ejections.

\begin{figure}
  \begin{center}
    \includegraphics[width=8.0cm]{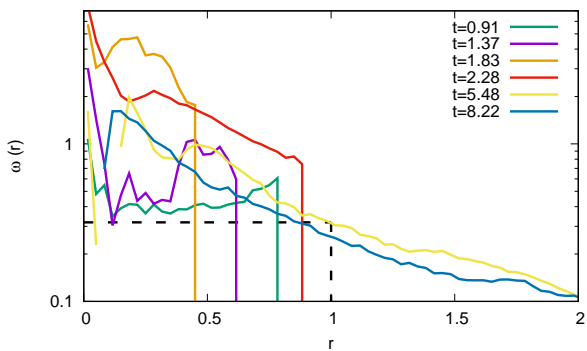}
  \end{center}
  \caption{The angular velocity as function of radius $\omega (r)$
    at indicated times for TC-B02. The vertical drop indicates the disk boundary.
    The initial profile is put in dashed line. The missing part of the
    line at the two last times indicates where $\omega$ is negative.
    This plot exhibits the development of the differential
    rotation after collapse, which is necessary to form the spiral arms.}
  \label{fig_cr_omega_begin}
\end{figure}

For further details on the winding process, the angular velocity as a
function of radius $\omega (r)$ at different times of TC-B02 is shown
in Fig. \ref{fig_cr_omega_begin}. We find that, while the disk is
shrinking from $0.91$ to $1.37 \ t_{d}$, the particles with high angular
velocities are developed near the disk boundary, forming the
peripheral ring observed in Fig. \ref{fig_circular_early}.
On the contrary, the rotational velocity of the inner part is slightly
developed, allowing the separated clumps to be formed by fluctuations.
In the subsequent time, the particles with high angular velocities
from the boundary finally fall to the center and the disk reaches
the minimum size. After the turning point, 
$\omega$ beyond $r\sim 0.5$ develops the
differential pattern because the particles reduce their angular speeds
while they are travelling outwards due to the conservation of 
angular momentum, shearing
the mass flow to become spiral arms in this process.
In the innermost part, i.e. below $r \sim 0.2$, the rotation
speed drops and becomes near zero or negative at the
last two plots. This can be that this part is dominated by the more
isotropic velocity distribution caused by the strongly varying potential
during the violent relaxation that re-aligns the particle velocities.

To summarize the formation process of the spiral pattern in a TC disk,
there are three necessary ingredients. These include the concentration
of mass flow, the persisting mass ejection and the differential rotation.
The formation steps can then be detailed as follows. First, while the
rotating disk collectively collapses, the motions of particles are
locally modulated, making the particle flows more concentrated in
some paths. Then, by a combination of the continual ejection and
the differential rotation due to the conservation of angular momentum,
the disk center acts as the rotating source of ejection
that feeds the spiral imprint of particles outwards.
The formation of multi-arm pattern has also been
reported for a disk in equilibrium. It has been shown that the
local over-densities embedded in the disk, either by the
Poissonian noise \citep{fujii_et_al_2011} or by the randomly 
placed giant molecular clouds \citep{donghia_et_al_2013},
play a similar role as in our case since they
can trigger the local concentrations, which
are then sheared by the differential rotation and become
the spiral arms.

\subsection{Dependences on $N$ and $b_{0}$} \label{n_b0_cr}

\begin{figure*}
  \begin{center}
    \begin{tabular}{ccc}
      \hspace{-8mm}
      \includegraphics[width=4.5cm]{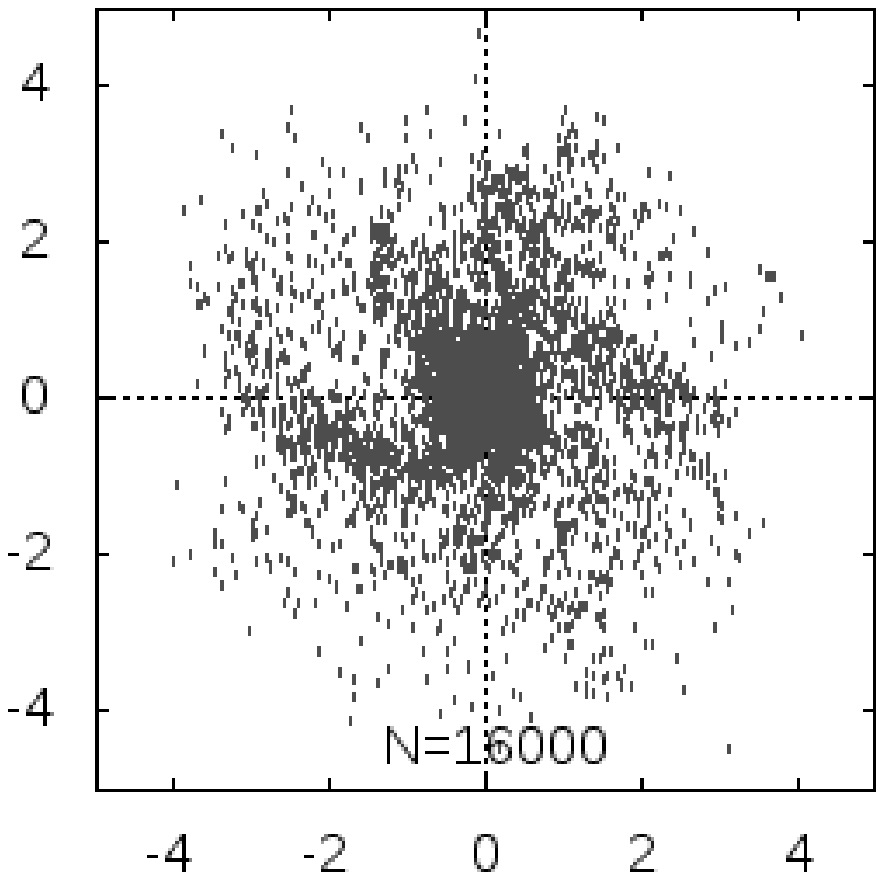} \hspace{-5mm}
      & \includegraphics[width=4.5cm]{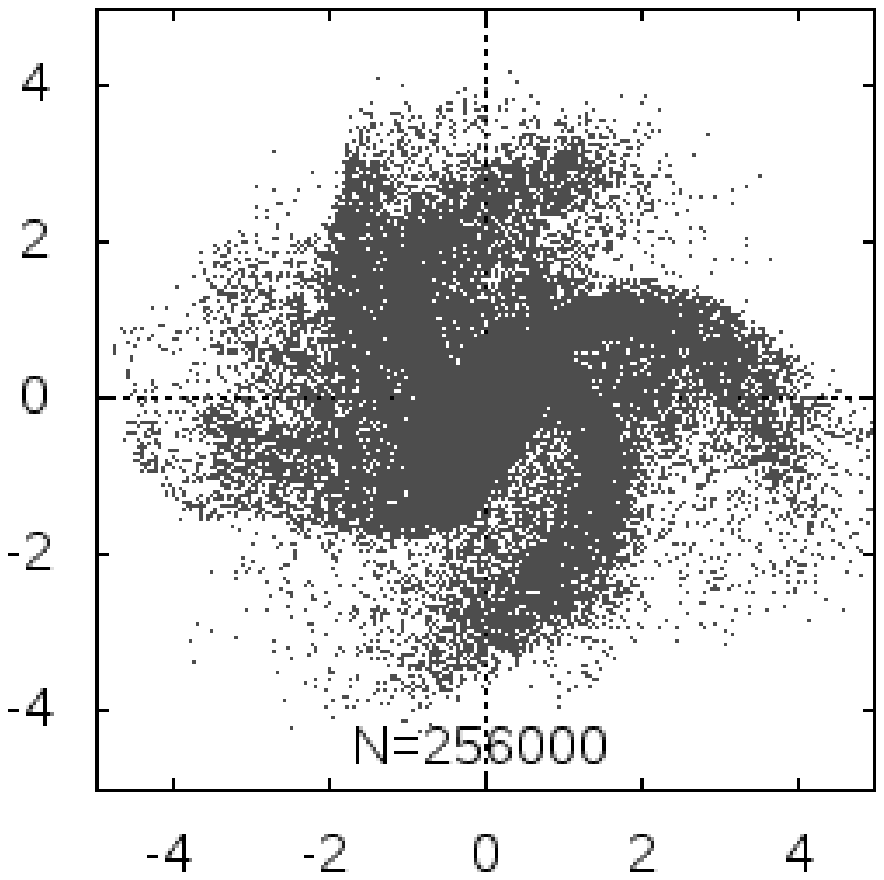} \hspace{-5mm}
      & \includegraphics[width=4.5cm]{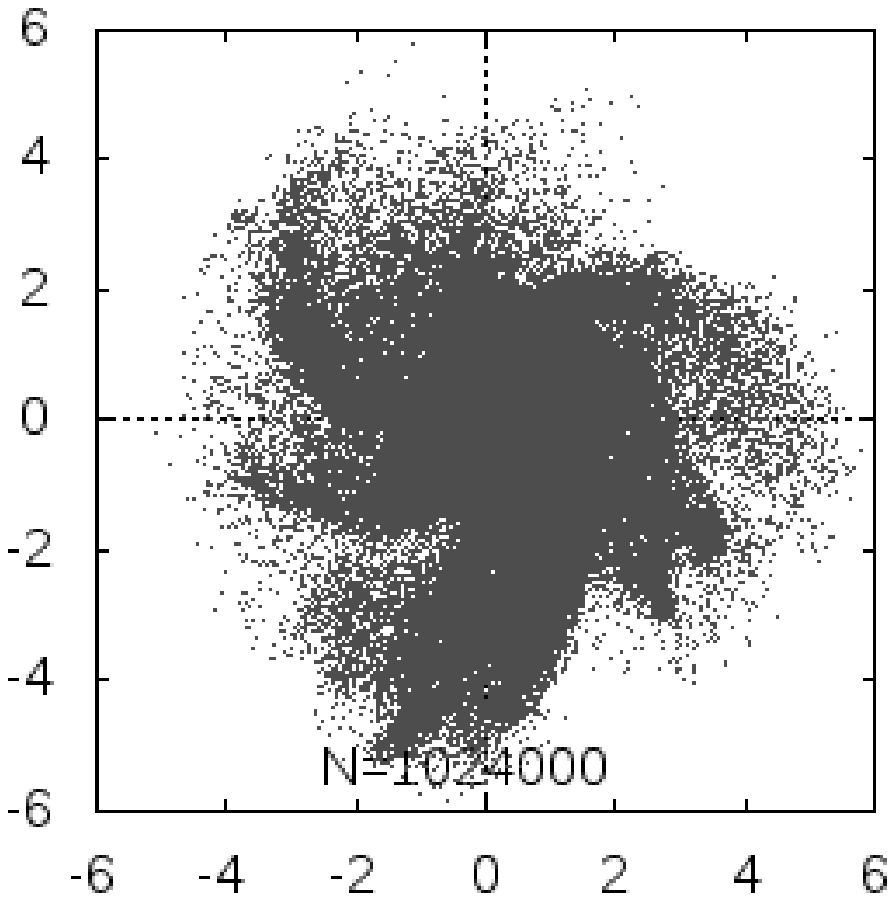} 
    \end{tabular}
    \caption{TC-B02 for $N=16,000$ and $256,000$ at $5.48 \ t_{d}$ 
      (left and middle panels); and $N=1,024,000$ at $6.39 \ t_{d}$ 
      (right panel). The number of arms increases
        with $N$, as speculated by our hypothesis.}
    \label{fig_test_n_cr}
  \end{center}
\end{figure*}

As demonstrated in Sec. \ref{early_winding}, the formation of
the spiral pattern in a TC disk relies on finite-$N$ fluctuations
in combination with the differential rotation and
ejection. We inspect further the dependence on $N$ and $b_{0}$
to the resulting arms in this disk family.
First, the configurations of TC-B02 with different $N$ are shown in 
Fig. \ref{fig_test_n_cr}. With $N$ covering almost $3$ orders of magnitude,
it becomes clear that more spiral arms are generated as we increase $N$. 
For $N=16,000$, we observe two opposite arms on the left and 
right of the nucleus. As $N$ goes up to $256,000$,
we observe four main arms with various length and thickness.
These arms emerge from different random positions. Thus, they are
originated from the separated flows of mass.  
With $N=1,024,000$, the spiral arms are even more tightly arranged.

To explain and make a simple estimate for this result, we recall
the procedure presented in Fig. \ref{fig_circular_early}. 
In order to establish the spiral arms, particles must be
accumulated before they are driven out by the ejection. We then
hypothesize that this process involves two competing collapse
time-scales since the start. The first is the global time-scale,
$\tau_{0}$, for the entire disk which can be obtained by the
dimensional approximation of the acceleration at disk boundary to be
\begin{equation}
  \tau_{0} \sim \sqrt{\frac{r_{d}}{G\Sigma_{0}}} \label{tau0}
\end{equation}
where $r_{d}$ is the disk radius and $\Sigma_{0}$ is the surface density
at start. In the same way, the local collapse time, $\tau'$, is given by
\begin{equation}
  \tau' \sim \sqrt{\frac{r'}{G\Sigma'}} \label{tau_prime}
\end{equation}
where $'$ designates the local variables.
By this definition, $r'$ corresponds to the effective radius of
the cloud, in which the density is equal to $\Sigma '$.
We then make an assumption that the particle number forming
each arm and its adjacent part is $N'$
that is equally divided by the final number of arms $n_{a}$.
We then have $N'=N/n_{a}$.
This proposition also implies that there are $n_{a}$ sub-collapses,
hosting $N'$ particles in each one, prior to the ejection. 
For $\Sigma'$, it is more convenient to express it as
$\Sigma'\equiv \Sigma_{0} +\Delta\Sigma'$ where $\Delta\Sigma'$
corresponds to the deviation from $\Sigma_{0}$ in the area initially
occupied by $N'$. Then, $\tau'$ can be re-written in a linear
approximation form as
\begin{equation}
  \tau' \equiv \tau'_{0}+\Delta\tau' \sim \sqrt{\frac{r'}{G\Sigma_{0}}}
  -\frac{1}{2}\sqrt{\frac{r'}{G\Sigma_{0}}}
  \Bigg( \frac{\Delta\Sigma'}{\Sigma_{0}}\Bigg). \label{delta_tau_prime}
\end{equation}
For an arm to be successfully built, our hypothesis is that 
$\tau' \ll \tau_{0}$ so that the concentrated
cloud is set well before the maximum collapse and ejection. 
This can be satisfied given that
\begin{equation}
  \tau'_{0}\sim |\Delta\tau'|. \label{compare_time}
\end{equation}
By some manipulation, the expression (\ref{compare_time}) yields
$\Delta\Sigma'\sim\Sigma_{0}$, i.e. the local density fluctuations are
of order the mean density. If the relative density is approximated by 
Poissonian noise as $\Delta\Sigma'/\Sigma_{0}\sim 1/\sqrt{N'}$, 
the condition (\ref{compare_time}) otherwise gives 
\begin{equation}
  n_{a} \propto N. \label{na_n}
\end{equation}
Equation (\ref{na_n}) qualitatively agrees with the result in the way
that we can have more spiral
arms if more particles are introduced. However, as evaluated from
Fig. \ref{fig_test_n_cr}, this expression evidently over-estimates the
simulated results: the number of arms for $N=1,024,000$ is not eight-fold
from the reference case. This might be that 
real process is not as simplistic as we initially thought.

\begin{figure}
  \begin{center}
    \begin{tabular}{cc}
      \hspace{-5mm}
      \includegraphics[width=4.9cm]{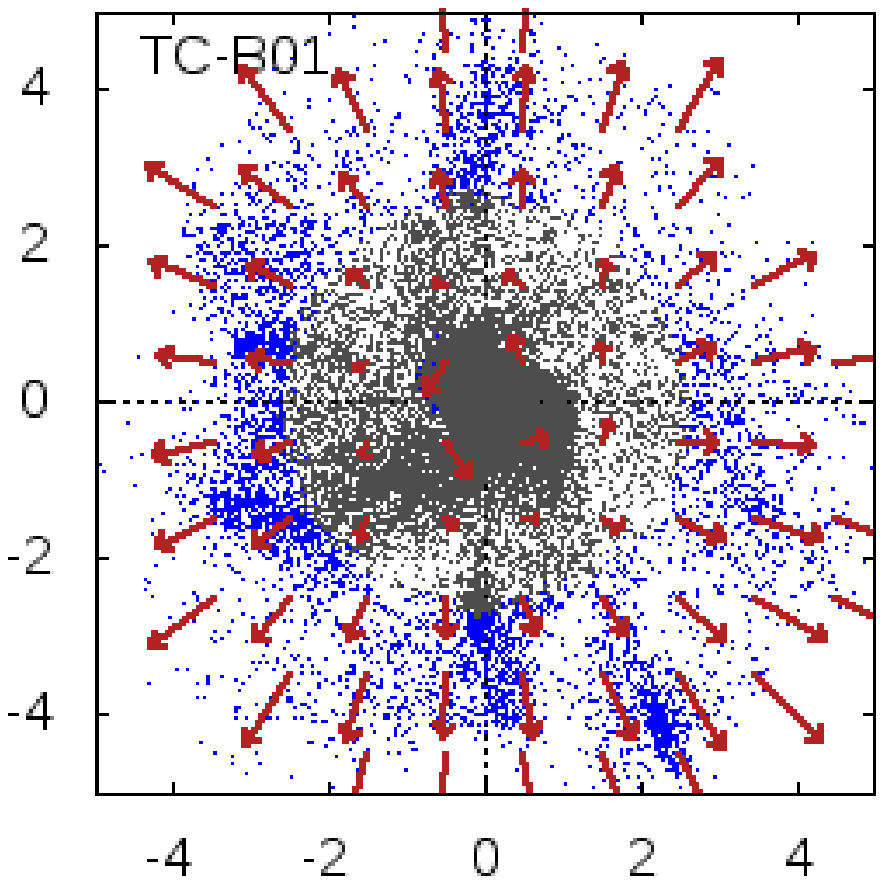} \hspace{-10mm}
      & \includegraphics[width=4.9cm]{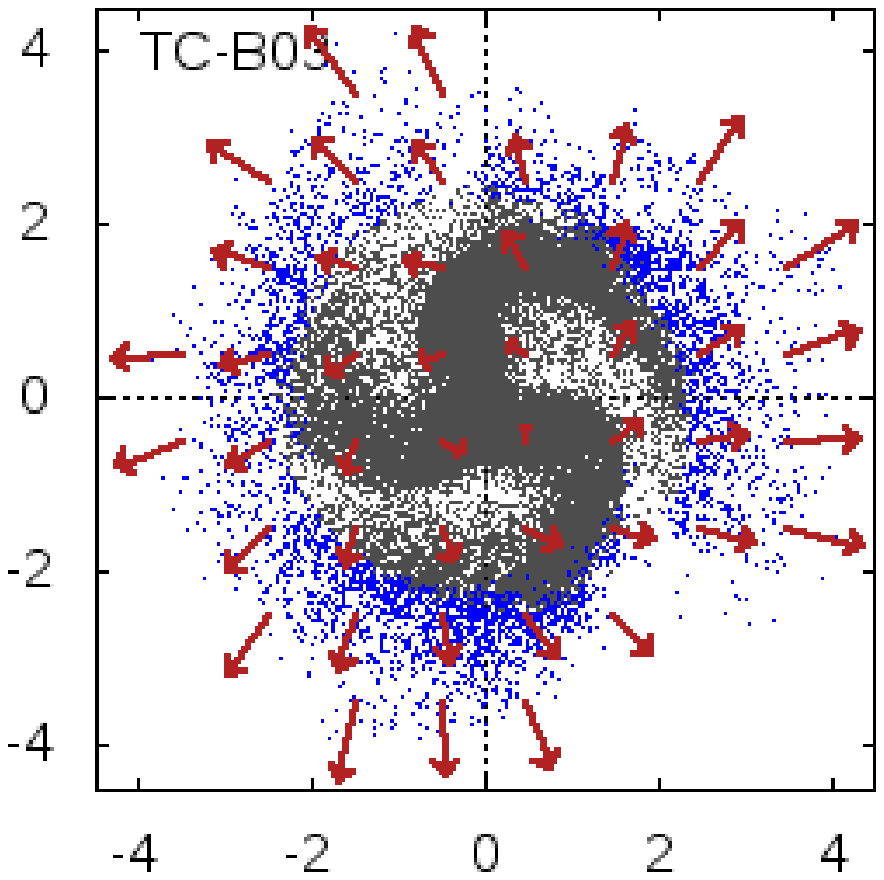} 
  \end{tabular}
  \end{center}
  \caption{The configurations of TC-B01 (left panel) and 
    TC-B03 (right panel) at $5.48 \ t_{d}$ with velocity fields superposed.
    Bound and unbound particles are put in different colors.
    The outer part of spiral arms consists
    of unbound particles moving outwards, while the inner bound
    part is dominated by the tangential motion. The spiral arms are
    winding more as we increase the virial ratio $b_{0}$.}
  \label{fig_last_circular}
\end{figure}

We then consider the spiral pattern with different $b_{0}$.
Shown in Fig. \ref{fig_last_circular} are the configurations of two other
TC at $5.48 \ t_{d}$ with the velocity field in an arbitrary unit.
First of all, we find that the outer part of the arms is composed of
the unbound particles, as expected from the primary ejection.
For B03, we observe the multi-arm pattern embedded in a diluted background
similar to TC-B02 and the arms are winding more around
the nucleus. With $b_{0}=0.1$, we capture instead the unwound
filaments pointing outwards from the nucleus.
Thus, it turns out that, in the latter case, the mass accumulation
can be accomplished but it fails to wind up.
The fact that the winding is enhanced if $b_{0}$ is increased from
$0.2$ to $0.3$ is in line with the approximation given
in equation (\ref{eq_mot_disk_dthetadotdr}). However, the
failure in producing the spiral arms at low $b_{0}$ is an unexpected
result as that approximation predicts rather less winding.
This could be that the shearing fuelled by the initial rotation
is not strong enough to withstand the violent relaxation and the
local mass accumulation in the early stage,
which tend to randomize the particle velocities.

For the detailed kinematics of the spiral structure,
the velocity field reveals characteristic motions at different parts.
We find that the nucleus and inner bound region are dominated
by the tangential motion in the direction of the initial rotation.
Further outwards, the motion of the outer bound region
is combined between radial and tangential components.
As we cross the boundary, the radial motion dominates the velocity
of unbound particles, as expected from the ejection.
They are travelling with far greater speed than those inside.
Despite that the bound and unbound parts are dynamically dissimilar,
we do not see the discontinuity between the two sides of the arms. 
The size of the disk and spiral arms
is considerably extended from the initial size.
We speculate that there are two distinct mechanisms  
responsible for the expansion of the disk.
The first mechanism is the violent relaxation that generates
an amount of ejected particles, constituting the outermost part
of the system. This mechanism has been described by 
\citet{joyce+marcos+sylos_labini_2009} that the late-arriving
particles gain the kinetic energy from the central
time-varying potential of the inside mass. 
Those that gain sufficiently high
energies then escape from the system.
We anticipate that this part will continually expand due
to the non-zero terminal velocity. Another possible mechanism
may be the radial migration (or radial mixing),
which comes into play when the spiral pattern is formed.
This mechanism describes the expansion of the disk outskirts
as a consequence of the angular momentum re-distribution by
the spiral disturbance \citep{sellwood+binney+2002, roskar_et_al_2008}.
Alternatively, the bar-spiral interaction \citep{minchev+famaey_2010}
or the passage of satellite \citep{quillen_et_al_2009, bird_et_al_2013}
has also been reported to cause the same migration.
By this mechanism, the particles with high angular momentum gain
can migrate outwards, while remaining in close orbits.
This explains why the boundary of the bound structure can be
further expanded although the disk is already in a virialized state.

\subsection{Spiral arms from TN disks: trial for grand-design structure} 
\label{spi_kine}

\begin{figure}
  \begin{center}
    \begin{tabular}{cc}
      \hspace{-6mm}
      \includegraphics[width=4.8cm]{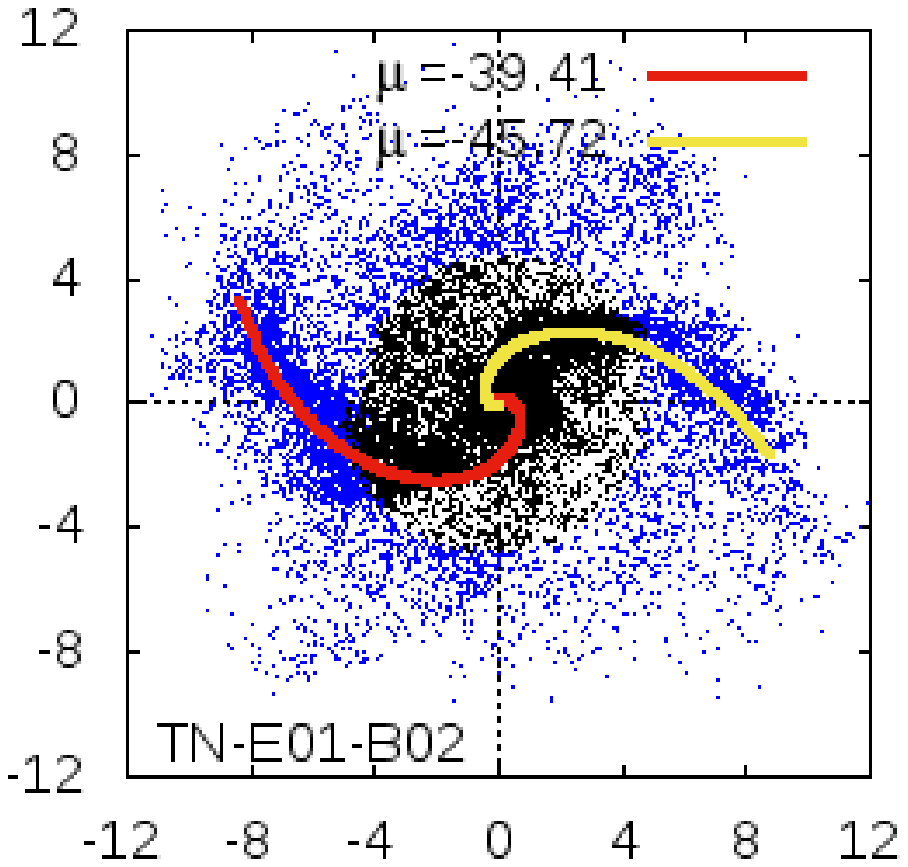} \hspace{-8mm}
      & \includegraphics[width=4.8cm]{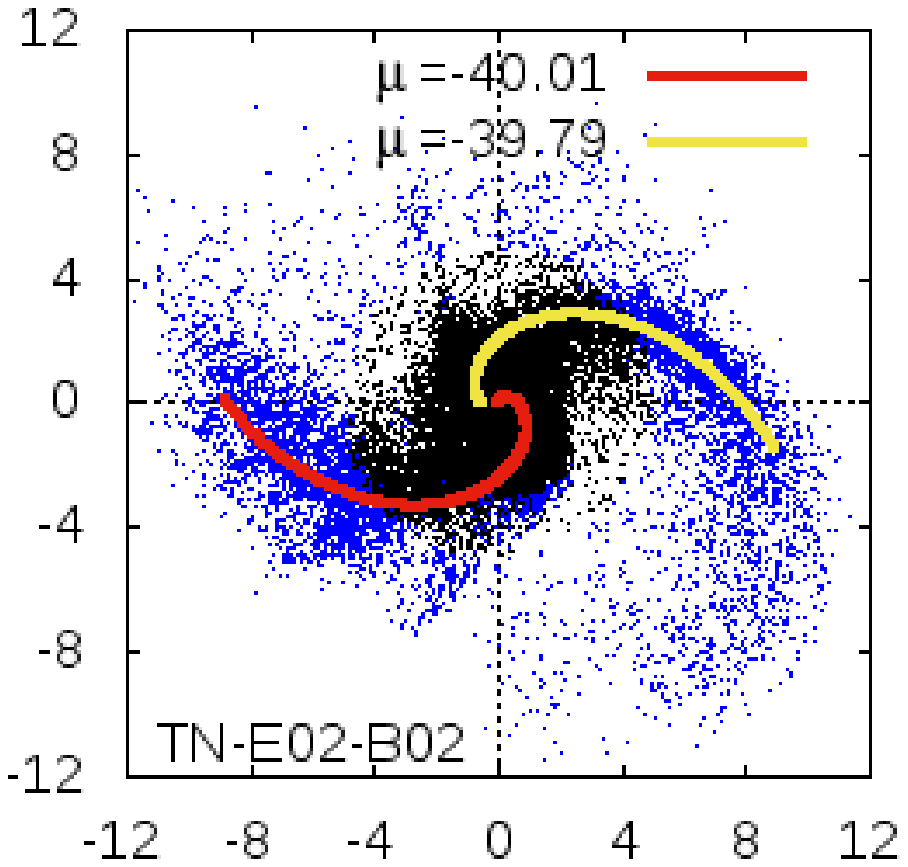} \\
      \hspace{-6mm}
      \includegraphics[width=4.8cm]{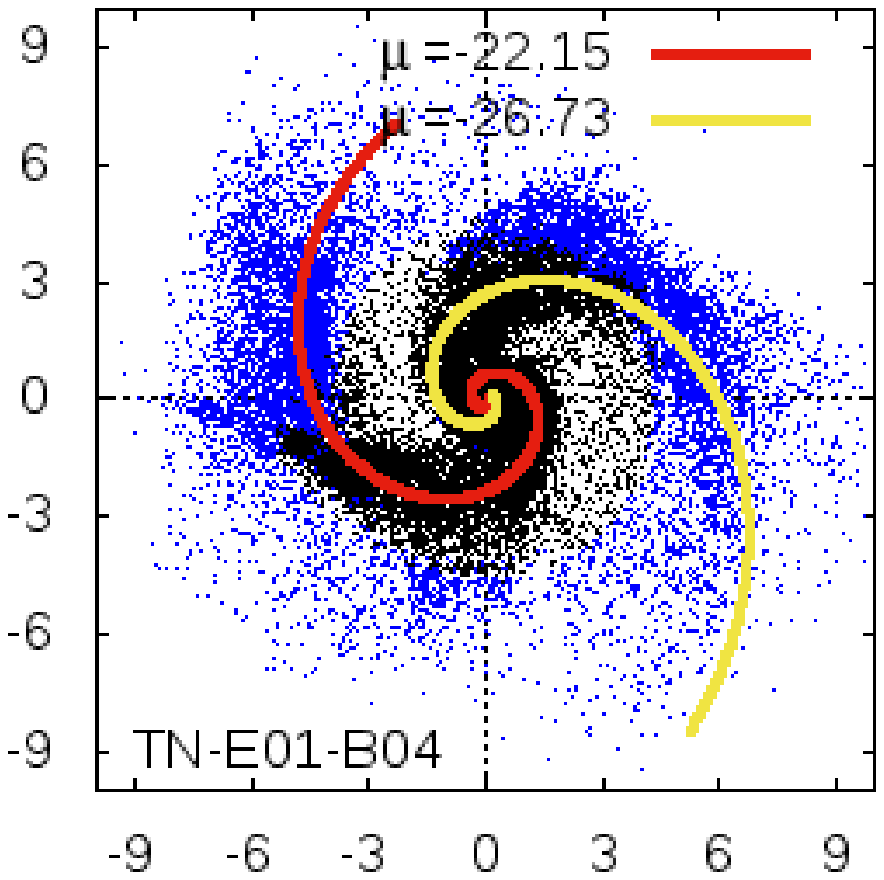} \hspace{-8mm}
      & \includegraphics[width=4.8cm]{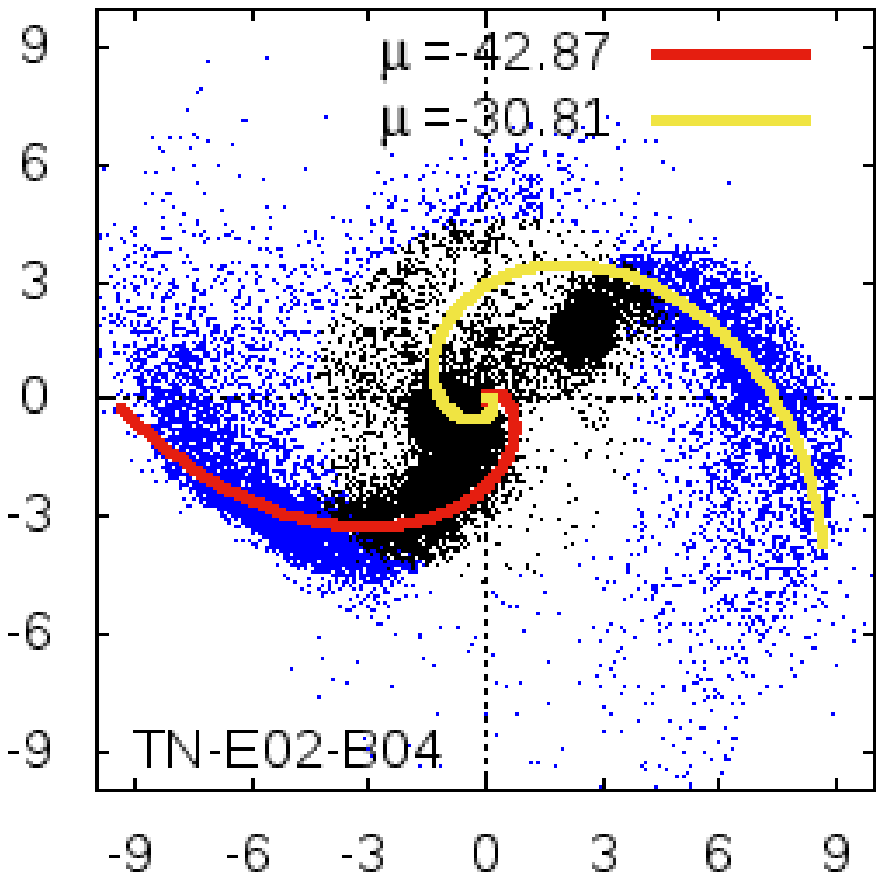} 
    \end{tabular}
  \end{center}
  \caption{The configurations of selected TN at $10.96 \ t_{d}$ with the
    best-fitting logarithmic spiral functions and the corresponding
    pitch angles $\mu$ for each arm. 
    Bound and unbound particles are distinguished by different colors.
    We observe two prominent arms emerging from the opposite 
    sides of the core, indicating the control of mass ejection by
    disk surface asymmetry.}
  \label{fig_pitch_arm}
\end{figure}

In this section, we investigate the formation of the spiral structure in TN family.
To verify this, some selected TN at $10.96 \ t_{d}$ are illustrated
in Fig. \ref{fig_pitch_arm} with bound and unbound particles distinguished.
The best-fitting spiral function (\ref{pitch_def}) and the pitch angle $\mu$
are also put in the figures. In all cases, we observe two spiral arms emerging
from the opposite sides, with unbound particles occupying more than half of
their lengths. Comparing with Fig. \ref{fig_last_circular}, we note that
the radial size of the bound mass is further expanded in the course of
a few dynamical times. This is another evidence of the radial migration
caused by the interaction with spiral arms.

To explain the emergence of the two controlled arms, let us recall
the relating work of \citet{benhaiem+sylos_labini_2015}. 
During the collapse of a non-rotating ellipsoid, they demonstrate that
the inertial principal axis of ejection is strongly aligned with
the initial major axis of the system. To explain this, one first recalls
the analysis of the non-isotropic collapse by \citet{lin+mestel+shu_1965}.
Their analysis informs that the particles at the furthest side
of the ellipsoid have the longest free-fall times.
As a consequence, they arrive at the center later and gain the kinetic
boost from the central field created by particles arriving earlier
to escape from the system. 
From our comprehension, that mechanism definitely plays a role in our case for
guiding the ejection through the major axis. We further
speculate that this alone cannot establish the visually detectable arms.
We may again underline
the necessity of the pre-ejection concentration as another important factor.
This localized mass accumulation can also be explained by an alternated 
view of the same theory by \citet{lin+mestel+shu_1965}.
The fact that the shortest axis has the shortest free-fall time lets it reach
the origin first while the infall from the longer side is still ongoing,
forcing the mass concentration to be out of the minor axis.
This explains why our collapsing ellipse 
develops two arms, supposedly along its initial major axis.
In addition, TN-E01-B01 has been tested with $N$ from $16,000-256,000$.
We find that the number and position of main arms can effectively be
controlled in that range of $N$. Thus, the modification of disk surface
is effective to subdue the finite-$N$
effect and confine the particle concentration and ejection.

We then inspect the agreement with the best-fitting logarithmic spiral
function. To determine $\mu$, we first calculate the surface
density of a point-wise configuration in a polar grid. Then, starting
from the radius where the arm begins, the cell with maximum density
residing in the visible arm is picked at each radius until the end of
the arm. Finally, $\mu$ is obtained by fitting the positions of chosen
cells with equation (\ref{pitch_def}). The fitting yields negative $\mu$ as
the arms are trailing and we will keep this sign convention for
further analysis. Without the separation of nucleus,
the constant-$\mu$ model fits well with the simulated configurations
resembling to the grand-design structures. Particular concern
is on TN-E02-B04 where we note the separation into two sub-nuclei, each
of which attaches to its individual arm.
We note slight deviations from the spiral lines of both arms due to
the displacement of sub-nuclei from the origin.
The separation of nucleus is possibly because
high $e_{0}$ leads to a more elongated shape at collapse.
Thus, it is more fragile to be torn apart by the rapid rotation.

\begin{figure}
  \begin{center}
    \includegraphics[width=7.5cm]{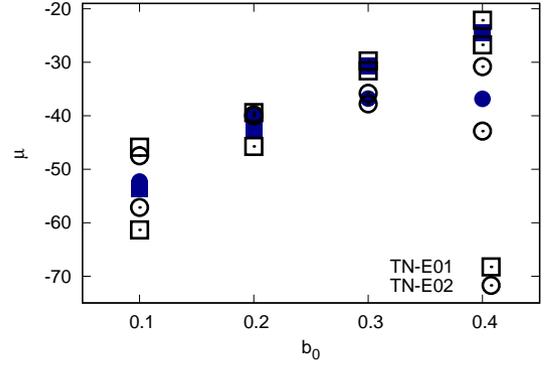}
  \end{center}
  \caption{The pitch angle $\mu$ as function of
    the initial virial ratio $b_{0}$ for each arm of TN-E01 and TN-E02 
    taken at $10.96 \ t_{d}$ in empty points. Averaged $\mu$ from both arms
    in each case is put in filled points with the same shapes.
    The increase of $\mu$ with $b_{0}$ is predicted by our analysis.}
  \label{fig_mu_b0}
\end{figure}

To see more quantitatively the variation of pitch angle,
the plot of $\mu$ of all TN disks at $10.96 \ t_{d}$
as function of $b_{0}$, two individual $\mu$ as well as the average
for each case, is shown in Fig. \ref{fig_mu_b0}. For both $e_{0}$,
we find that $\mu$ increases accordingly with $b_{0}$ until $b_{0}=0.3$,
i.e. the arms are more tightly wound.
That tendency is retained for E01 as we progress to B04 while for
E02, the winding is apparently limited from advancing higher.
The dispersion of $\mu$ between the two arms is typically less than
$15^{\circ}$ and apparently does not correlate with $e_{0}$ or $b_{0}$.
The fact that $\mu$ increases with $b_{0}$ is predicted by
our estimate (\ref{eq_mot_disk_dthetadotdr}) which implies that
the shearing strength increases with the initial rotation.

Let us compare the situation between TC and TN apart from the
ability to control the arm position and number.
The winding dynamics in both families is similar in the way that 
it tends to wind up more as we increase $b_{0}$.
However, this tendency is apparently limited beyond TN-E02-B03.
At the lower end, the winding does not take place
at TC-B01 but it does for its TN counterparts. This can be explained
by that, with a guided ejection, the collective flows of mass are more
intensified. Therefore, their initial motion is hardly overcome
by the disturbance during the violent relaxation
and is preserved until the ejection.
More discussion relating to the effect from the disk configuration
will be available in Sec. \ref{spi_thick}.

\subsection{Long-time behavior of spiral arms} \label{long_time}

\begin{figure}
  \begin{center}
    \begin{tabular}{cc}
      \hspace{-6mm}
      \includegraphics[width=4.8cm]{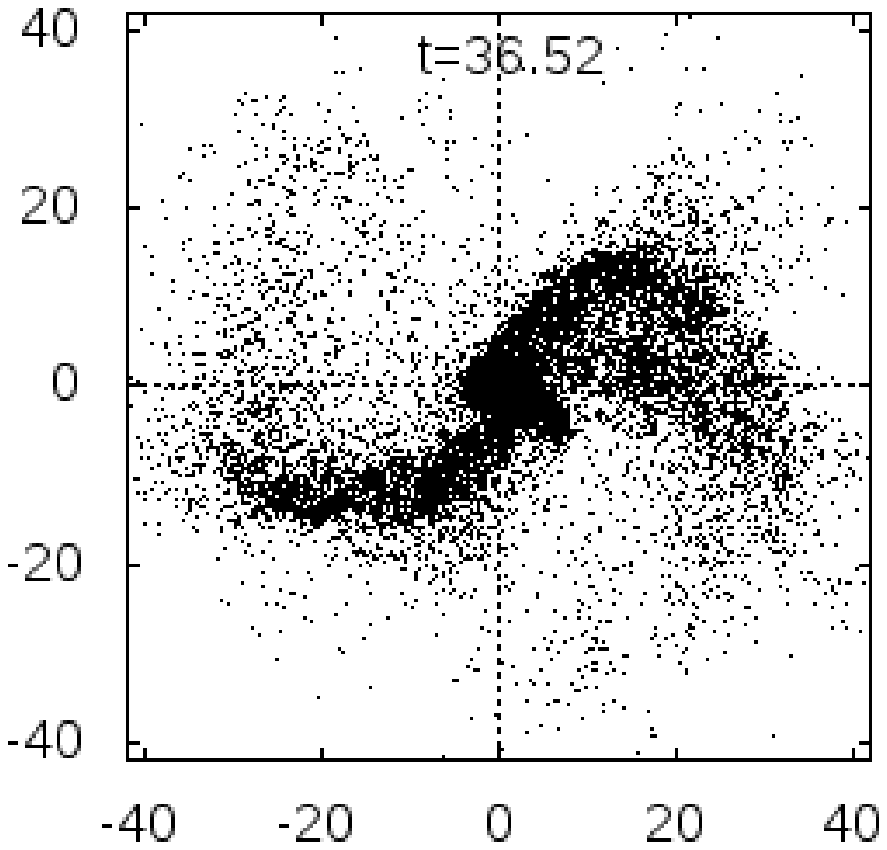} \hspace{-7mm}
      & \includegraphics[width=4.8cm]{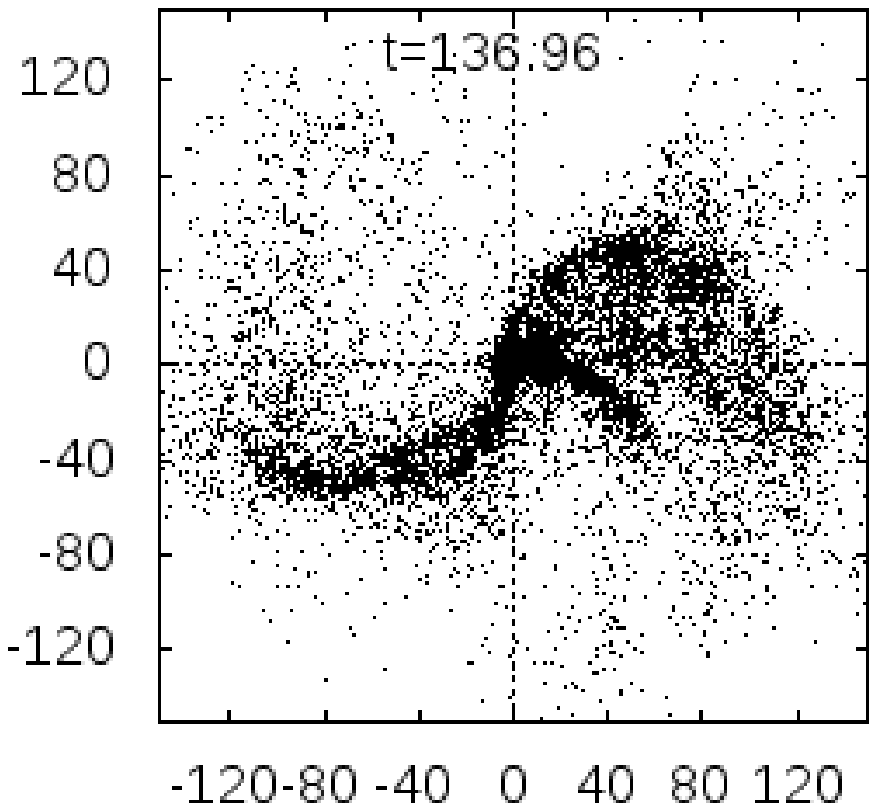}  
      \hspace{-5mm}
    \end{tabular}
  \end{center}
  \caption{The projection of prolonged TN-E01-B01 at indicated times.
    It appears that the spiral structure obtained
    in this way is a robust structure that, despite the
    continual outward motion, remains in shape
    for times much longer than the formation time-scale.}
  \label{fig_thin_arm_long_1}
\end{figure}

In this section, we study the long-term behavior of the spiral arms.
Some TN cases are evolved further and the configuration of the
prolonged TN-E01-B01 until $136.96 \ t_{d}$
is shown in Fig. \ref{fig_thin_arm_long_1}. First of all, we find
that the obtained spiral structure is robust: it remains in shape
for time much longer than the formation time-scale
without any sign of dissolution or disfigurement. Its size is much
expanded from that in Fig. \ref{fig_pitch_arm} as a result of the
radially outward motion. We observe an additional morphological change
of the nucleus at $36.52 \ t_{d}$ where another
short arm emerges and stretches out rapidly as observed at $136.96 \ t_{d}$.
Apart from that fast emergence, there is no further change of the
main arms. The overall configuration is almost self-similar by
the radial stretching. The rotation is weakened
so it is hardly detectable by visual inspection.
This might be that the particles of the spiral arms are approaching their
terminal velocities which rather follow the linear trajectories.

\begin{figure}
  \begin{center}
    \begin{tabular}{c}
      \includegraphics[width=6.2cm]{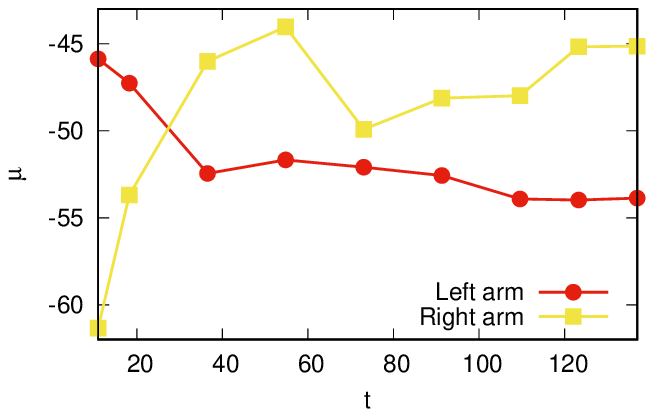} \\
      \includegraphics[width=6.2cm]{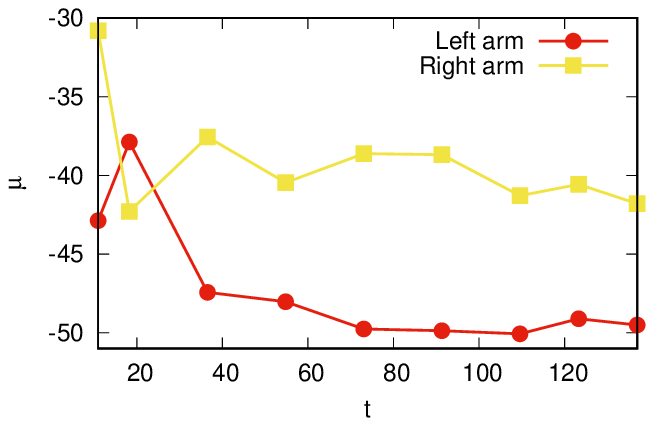}
    \end{tabular}
  \end{center}
  \caption{The temporal evolution of the pitch angle
    $\mu$ for each spiral arm of the prolonged
    TN-E01-B01 (top panel) and TN-E02-B04 (bottom panel).
    It is evident that the radially outward motion tends to dominate
    the long-term evolution by means of decreasing $\mu$.}
  \label{fig_t_mu}
\end{figure}

To examine the evolution of spiral arms more quantitatively,
the plots of the temporal evolution of $\mu$ for each arm of
the two most
prolonged cases are depicted in Fig. \ref{fig_t_mu}. We observe the
same behavior for three of four arms where $\mu$ evolves quickly
(either increasing or decreasing) until $\sim 40 \ t_{d}$. Then, it 
declines modestly by a few degrees over
a time interval about $90 \ t_{d}$.
The evolution of the right arm of TN-E01-B01 is different as $\mu$
does not mark a clear increase or decrease after its sharp increase
prior to $60 \ t_{d}$. To explain the temporal evolution of $\mu$,
we first recall the expression of $\mu$ from
equation (\ref{pitch_def}) and re-arrange it to be
\begin{equation}
  \tan \mu = \frac{1}{r}\bigg(\frac{dr}{d\phi}\bigg). \label{mu_gen}
\end{equation}
We then apply the time derivative to equation (\ref{mu_gen}) and,
after some manipulation, we obtain
\begin{equation}
  \dot{\mu}  = \frac{\sin (2\mu)}{2}\bigg( \frac{d}{dt}
  \bigg[ \ln\bigg(\frac{\dot{r}}{r\dot{\phi}}\bigg)\bigg]\bigg) \label{mu_dot}
\end{equation}
where $\cdot$ over the parameters indicates their time derivatives.
From this expression, the sign of $\dot{\mu}$ relies on
the fraction inside the logarithm which can be seen as the ratio 
of the radial to tangential velocity of arm element. With
$\mu\in [-90^{\circ},0^{\circ}]$, we justify that in radially and
tangentially dominated evolution, the pitch angle
is decreasing and increasing, respectively. Thus, the evolution
of the three regular arms in Fig. \ref{fig_t_mu} can be separated
into two different phases. First stage is the rapid variation 
due to the radial and tangential velocities competing 
to each other, allowing the sign of $\dot{\mu}$ to be either
positive or negative. Then, when the rotational velocity mostly
decays out by expansion, the motion is predominantly radial which
is seen by the gentle decline of $\mu$. 
The peculiar behavior of one irregular arm could be the result
from another arm emerging rapidly beside that potentially
disturbs the rectilinear and angular motions.

\begin{figure}
  \begin{center}
    \begin{tabular}{c}
      \includegraphics[width=7.5cm]{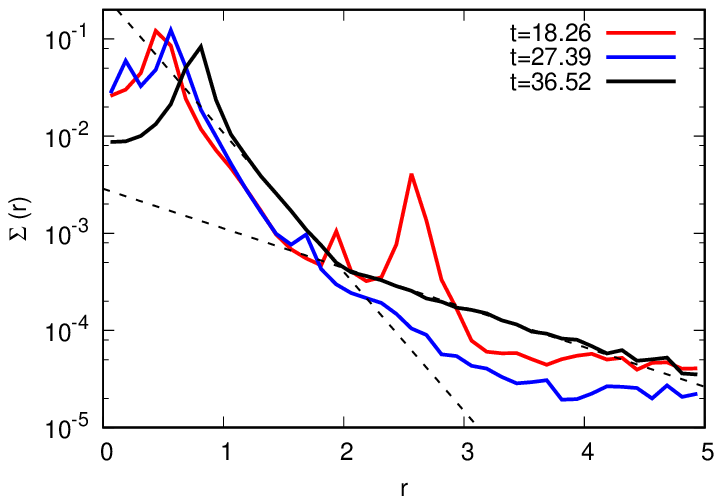} \\
      \includegraphics[width=7.5cm]{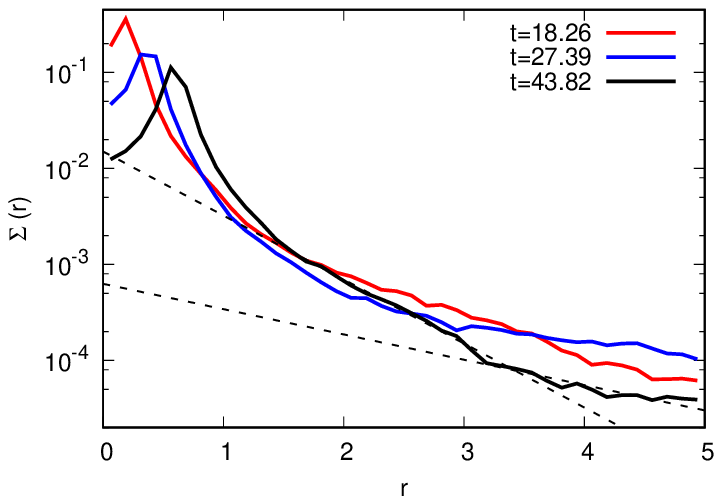}
    \end{tabular}
  \end{center}
  \caption{The surface density as function of radius $\Sigma (r)$
      for TN-E01-B02 (top panel) and TN-E02-B03 (bottom panel)
      at indicated times. In each panel, the two dashed lines
      correspond to the best-fitting exponential functions, performed
      with the profiles at the last time, for two different parts.
      This corresponds to the type-III profile.}
  \label{fig_density_long}
\end{figure}

From Fig. \ref{fig_t_mu}, it becomes clear that
the radial motion of the unbound part, which is originated from 
the gravitational collapse, dominates the long-term evolution
of the spiral arms. We verify further the particle distribution at
the inner part. Plotted in Fig. \ref{fig_density_long} is
the surface density as function of radius at different times
for two TN disks. For the profile at the
last time, we provide the best-fitting exponential functions 
for two different parts of the disk. The break radii corresponds
to the intersections of the two lines.
We find that the surface densities of both cases
are similar to the type-III (or anti-truncated) disk, which
is characterized by a slower exponential drop of the
density beyond the break radius. The exponential decay is
retained until $r \sim 5$, which is 5 times larger than
the initial size.
This indicates that the radial migration is the key mechanism
for expanding the boundary of the disk. The cause of
the type-III density profile is conjectured to be the gas accretion
into the disk outskirts, which enhances the velocity dispersion
in this region \citep{minchev_et_al_2012}.
Although our system does not involve
any accretion mechanism, we believe that the significant
random motion in the disk outskirts could be the remnant of
the gravitational collapse, which tends to randomize
the initial rotation. This explains why the  
disk relaxing in this way produces the type-III profile.

\subsection{Effect from increasing the thickness} \label{spi_thick}

\begin{figure}
  \begin{center}
    \begin{tabular}{cc}
      \hspace{-6mm}
      \includegraphics[width=4.5cm]{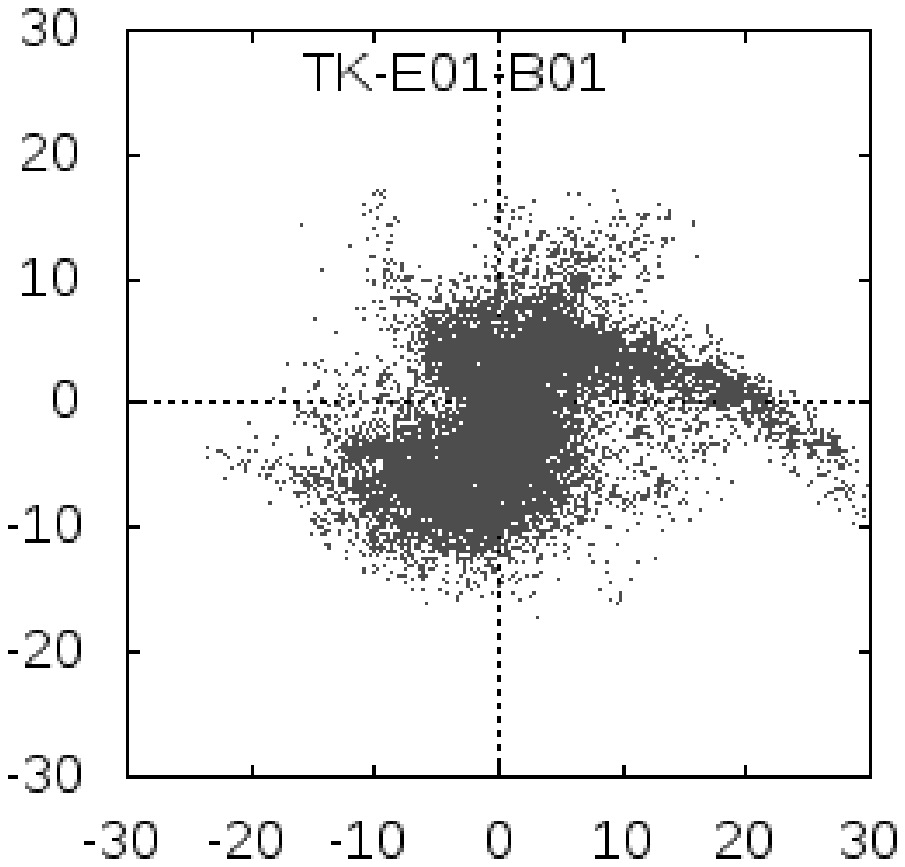} \hspace{-8mm}
      & \includegraphics[width=4.5cm]{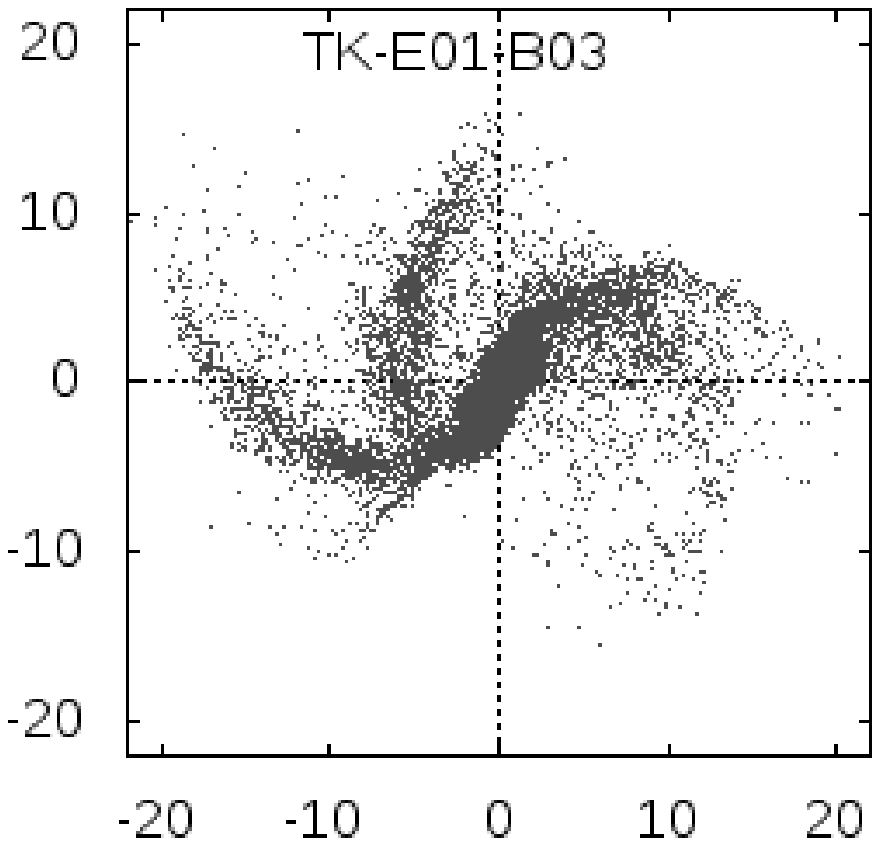} \\
      \hspace{-6mm}
      \includegraphics[width=4.5cm]{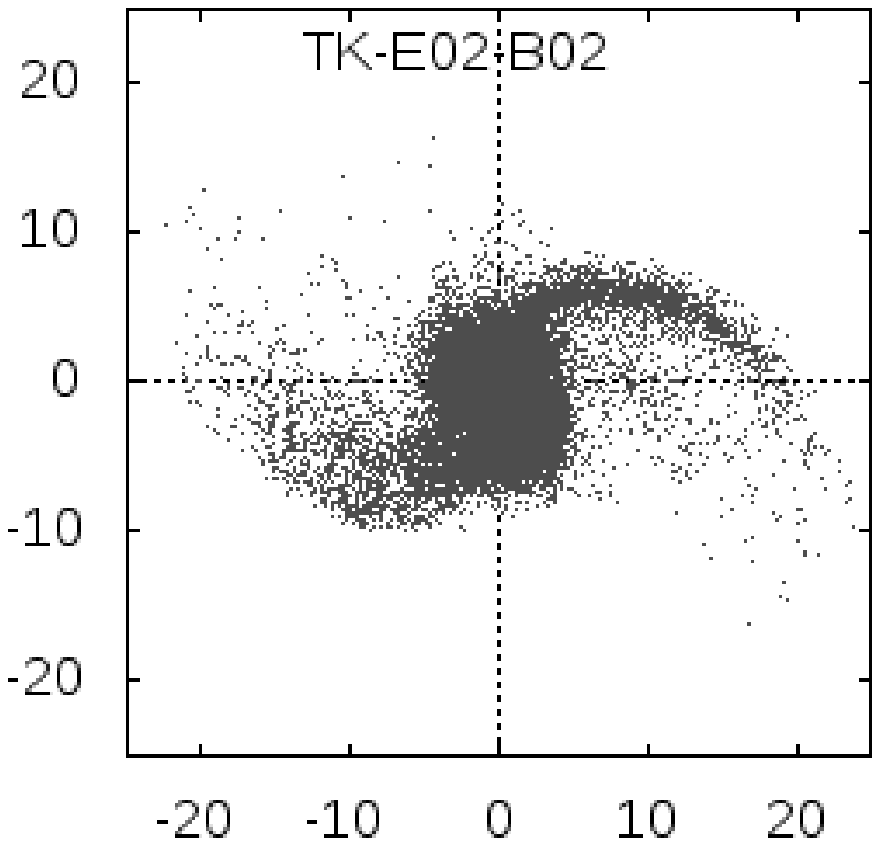} \hspace{-8mm}
      & \includegraphics[width=4.5cm]{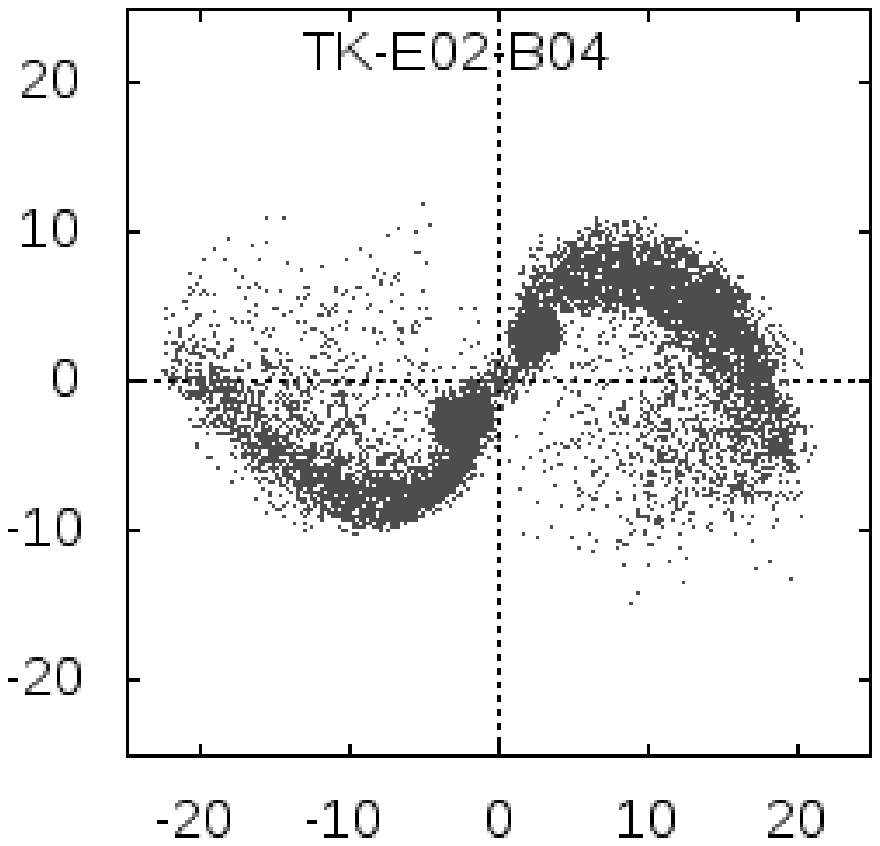}
    \end{tabular}
    \caption{The configurations of selected TK disks at $10.96 \ t_{d}$.
      The spiral patterns are more diversified
      than those formed in the TN cases.}
    \label{fig_thick}
  \end{center}
\end{figure}

We finally consider the spiral arms in the TK family. The projections
of some selected TK at $10.96 \ t_{d}$ are shown in Fig. \ref{fig_thick}.
Similar to the TN disks, we capture the trailing spiral structures but
their appearances are more diversified. We remark the imbalanced arms for
TK-E01-B01 and the additional third arm on the 
left side for TK-E01-B03. When $e_{0}$ reaches $0.2$,
overall appearances do not differ much from their TN counterparts.
The nucleus of TK-E02-B04 is also separated apart
similar to its corresponding TN.
In other words, the increase of thickness does not affect much the
final forms provided that $e_{0}\sim 0.2$.

\begin{figure}
  \begin{center}
    \includegraphics[width=7.5cm]{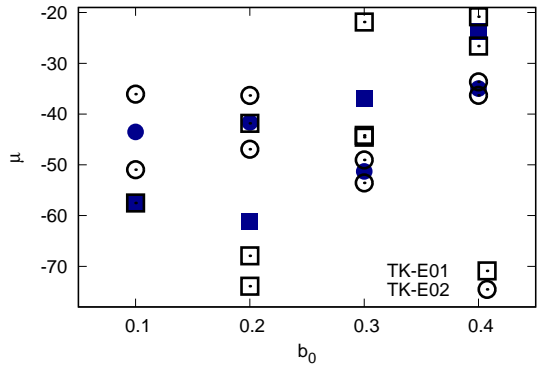}
  \end{center}
  \caption{The pitch angle $\mu$ as function of
    the virial ratio $b_{0}$ for TK-E01 and TK-E02 at $10.96 \ t_{d}$.
    The values of each individual arm and the averages are put in empty and
    filled points, respectively. Note that there are one point for 
    TK-E01-B01 and three points for each of TK-E01-B02 and TK-E01-B03.
    As the disk becomes thicker, the increase of $\mu$ with
    $b_{0}$ is less evident than it does in the plot for TN disks.}
  \label{fig_mu_b0_tk}
\end{figure}

The values of $\mu$ of each arm and the averages for
all TK cases are summarized in Fig. \ref{fig_mu_b0_tk}.
In the plot, there are one single $\mu$ for TK-E01-B01 and three 
for each of TK-E01-B02 and TK-E01-B03. The pitch angles of the
extra arms in the two latter cases are at the top-most points. 
For E01, the averaged $\mu$ for low $b_{0}$ is considerably lower than
that of the corresponding TN. The third arms of both cases yield 
much greater $\mu$ than the pitch angles of normal arms.
However, we still observe the 
increasing tendency of the averaged $\mu$ from B02 on, whether
the third arms are incorporated or not. Particular interest is on E02.
While the pitch angles are comparable to those from TN in
Fig. \ref{fig_mu_b0} and their dispersions are relatively low, 
the increase of $\mu$ with $b_{0}$ is not evident. 
They are rather fluctuating around $-40^{\circ}$.

\begin{figure}
  \begin{center}
    \begin{tabular}{c}
      \includegraphics[width=6.5cm]{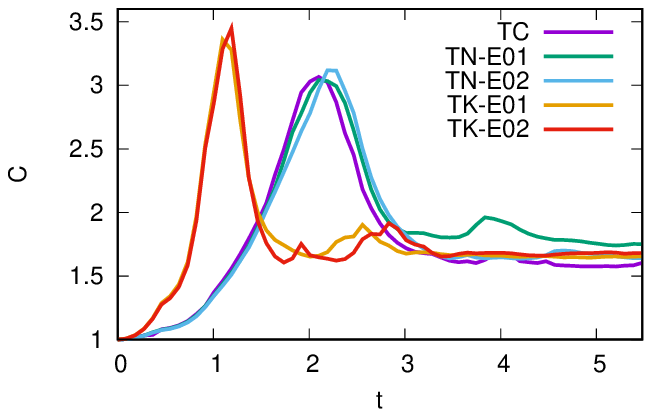} \\
      \includegraphics[width=6.5cm]{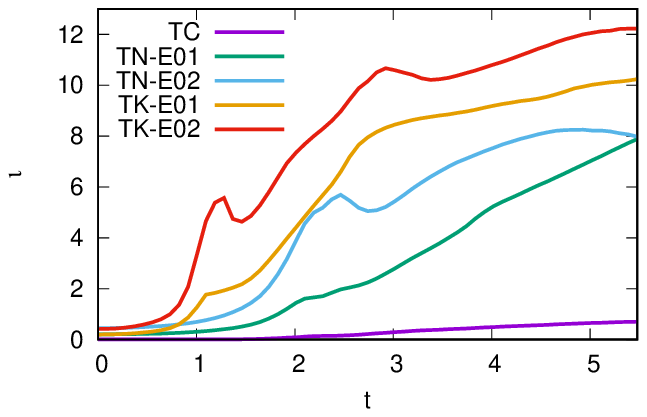}
    \end{tabular}
  \end{center}
  \caption{The temporal evolutions of the collapse
    factor $\mathcal{C}$ (top panel) and the flattening
    $\iota$ (bottom panel) calculated from bound particles for all
    B03 disks. The detail of calculation is given in text.
    These two plots suggest that the collapse of TK disks
    is more concentrated and more elongated than their TN counterparts
    that leads to some inconsistencies with the TN family.}
  \label{fig_rg_iota}
\end{figure}

In order to understand the inconsistencies with TN disks, e.g. the
diversity of the spiral structures and the non-increasing $\mu$ with $b_{0}$,
we will examine more parameters during the collapse and ejection. 
Shown in Fig. \ref{fig_rg_iota} are the temporal evolutions
of the collapse factor $\mathcal{C}$ and the flattening $\iota$
(see definitions in equations (\ref{col_def}) and
(\ref{iota_def})) calculated from bound particles of all B03 disks.
We find that all $\mathcal{C}$ are evolving in the same way:
it increases from $1$ before it relaxes down to a certain value
around $1.7$. The two TK disks evidently yield the maximum
$\mathcal{C}$ greater than those attained by the three thin cases.
It appears that, with more disk thickness,
the infall from third axis makes the contraction more compact. 
For the evolution of $\iota$, it remains low until the end
for TC. Thus, we may infer that the non-circular
effect is negligible here. The situation is
different for TN and TK where $\iota$ rises in accordance with
$\mathcal{C}$ until it reaches the maximum.
Afterwards, it is still increasing at a slower rate.
The values of $\iota$ at the maximum contraction of
TN and TK are almost equal but, further from that time,
the TK disks develop
higher flattening than that of TN until the end of plot.

These results suggest that the TK disks become more compact and
more elongated at the collapse than their TN counterparts.
We speculate that these are the causes of the inconsistencies
with TN disks as we will explain as follows.
Around the maximum contraction, the highly
non-axisymmetric field can potentially alter the motion of
particles which are arranged by the collective rotation beforehand.
Therefore, the effect of the initial rotation is washed away as seen
in Fig. \ref{fig_mu_b0_tk} where $\mu$ struggles or fails to
increase with $b_{0}$.
The additional effect from a highly asymmetric field is also seen
in TN-E02 and TK since we observe that the particles gather
closer to the arms than in TN-E01 and TC, in which they diffuse
throughout the inter-arm region (see Fig. \ref{fig_pitch_arm}). 
This is another indication of the forcefully guided ejection
by the self-consistent potential in TN-E02 and TK. 
To resolve the cause of the diversity of spiral patterns,
we suspect that it is the deeper potential field at collapse,
indicated by the maximum $\mathcal{C}$, that splits
the concentrations apart before the ejection, leading to
the separated arms on the same side of nucleus. However, this
occurrence is not observed for TK-E02 because stronger asymmetric
field created by higher $e_{0}$ can confine the mass more effectively
on each side thus it is more resistant to the central potential.

\section{Conclusion and discussion} \label{discussion}

In this work, we explore the spontaneous formation of the
spiral structure by gravitational collapse of a rigidly rotating disk.
From the start, the disk collapses and expands differentially so that
the mass outflow with density amplified is sheared and
forms finally the spiral pattern when looking on the disk plane.
Although trailing spiral arms are generic for all disk families,
a careful consideration suggests that their properties are different.
In TC disks, the absence of surface asymmetry leads to the formation
of multi-arm pattern regulated by the Poissonian noise.
In the formation process, the particles forming each arm are separately
grouping by local over-densities
before they are ejected out. This mechanism, in turn, leads to
the spiral structure with the number of arms that increases with the
particle number. For TN family, a slight modification on disk surface  
can confine the clustering along the longer axis of the elliptical
disk and guide the ejection through the same axis.
This yields the final structure 
with two prominent spiral arms on the opposite sides of nucleus similar
to the grand-design configuration. In these two families,
the pitch angles of the spiral arms tend to
vary in accordance with the degree of initial rotation.
When we increase the thickness, corresponding to the TK group, the situation
is more difficult as the relaxation leads to more visually complicated
structures or it even dissociates the winding degree from the initial rotation
in some condition. We suspect that these discrepancies are caused by a
deep non-axisymmetric central potential at collapse that splits apart
the mass concentration and intervenes the pattern of the pre-collapse motion.

About the kinematics of simulated arms, they are non-stationary as the
outer unbound part is continually moving outwards. The inner bound part
that consists of the virialized nucleus and the beginning of arms is, 
on the contrary, dominated by the tangential motion.
The surface density of the disk at the vicinity of 
nucleus also matches with the type-III profile. This is an evidence of 
the radial migration that arranges the disk outskirts.
Despite the continual expansion, these arms are 
robust as they remains intact for a long period of time.
The expansion affects only by decreasing slowly the pitch angle.
These characteristic motion and longevity are different from those of
the arms in real galaxies that are dominated by the differential rotation
and it is believed that they will dissolve after a number of rotations.
Although the properties of our spiral arms are incompatible with
the observed arms, the fact that this kind of spiral structure
(and also those in \citet{benhaiem_et_al_2017}) is obtainable
through the violent gravitational collapse
lets us speculate that this type of spiral structure
could arise in the realm where the galaxy relaxes straight
from a monolithic collapse. For an unstable protogalatic cloud
that possesses some amount of rotation, our mechanism suggests
that the spiral structure can potentially emerge out. 
The analysis of a cloud collapse in the cosmological framework
suggests that the relaxation of an overdense cloud should be
terminated and becomes galaxy around $z\sim 10-30$
\citep{partridge+peebles_1967a}. Thus, if the proposed
spiral structure exists, they might reside in
that range of redshift. To pinpoint them, the characteristic
velocity field is a useful identification tool.

\section*{Acknowledgements}
This project is supported by \fundingAgency{The Institute for
  the Promotion of Teaching Science and Technology (IPST)}
via the Research Fund for DPST Graduate with
First Placement (in fiscal year 2559 BE
with contract number \fundingNumber{003/2559})
under the mentorship of Khamphee Karwan. Numerical simulations
are facilitated by HPC resources of Chalawan cluster of the
National Astronomical Research Institute of Thailand (NARIT).

\end{document}